\documentclass[twocolumn,preprintnumbers,amsmath,amssymb,ieee,10pt]{revtex4}

\usepackage[english]{babel}
\usepackage[dvips]{graphicx}
\usepackage[T1]{fontenc}
\usepackage{makeidx}
\usepackage{dcolumn}
\usepackage{bm}

\begin{document}

\title{High Resolution Frequency Standard Dissemination via Optical Fibre Metropolitan Network}

\author{F. Narbonneau}
\author{M. Lours}
\author{S. Bize}
\author{A. Clairon}
\author{G. Santarelli}
    \affiliation{LNE-SYRTE, Observatoire de Paris, 61 Avenue de l'Observatoire, 75014 Paris, France}
\author{O. Lopez}
\author{Ch. Daussy}
\author{A. Amy-Klein}
\author{Ch. Chardonnet}
    \affiliation{Laboratoire de Physique des Lasers, Université Paris XIII, Villetaneuse, France}

\begin{abstract}

We present in this paper results on a new dissemination system of
ultra-stable reference signal at 100 MHz on a standard fibre
network. The 100 MHz signal is simply transferred by amplitude
modulation of an optical carrier. Two different approaches for
compensating the noise introduced by the link have been
implemented. The limits of the two systems are analyzed and
several solution suggested in order to improve the frequency
stability and to further extend the distribution distance.
Nevertheless, our system is a good tool for the best cold atom
fountains comparison between laboratories, up to 100 km, with a
relative frequency resolution of 10$^{-14}$ at one second
integration time and 10$^{-17}$ for one day of measurement. The
distribution system may be upgraded to fulfill the stringent
distribution requirements for the future optical clocks.

\end{abstract}

\maketitle

\section{Introduction}

\begin{figure}[!hbt]
  \centering
  \includegraphics[width=\columnwidth]{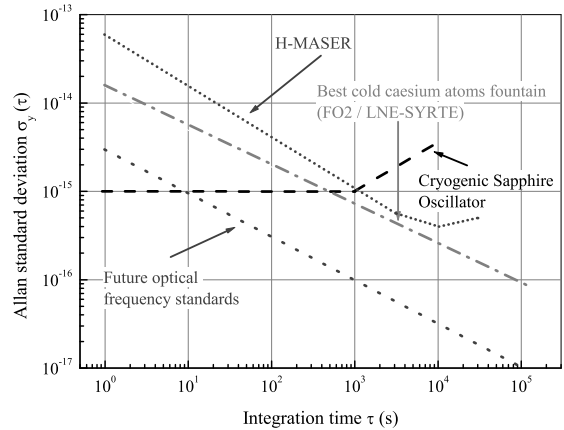}\\
  \caption{Allan deviation of some frequency standards}\label{specification_100MHz.EPS}
\end{figure}

Ultra-stable frequency and time sources play an important role in
many modern Time and Frequency metrology and fundamental physics
applications (clock evaluation, relativity tests, fundamental
constants test ...)(e.g. \cite{FundamentalTests},
\cite{Bizonours}, \cite{FISCHER}, \cite{PEIK}). In the field of
particles physics, modern large linear accelerators require RF
distribution system with minimal phase drifts and errors for the
neutrons and positrons generation \cite{refSLAC}. In
radio-astronomy, e.g. in the case of the ALMA (Atacama Large
Millimetric Array) project or for VLBI (Very Long Baseline
Interferometry), the combination of high frequency and long
baselines of the interferometer needs the distribution of a local
oscillator with low phase noise and low phase drift through the
array \cite{ALMA}, \cite{SATO2}. For the Deep Space Network (DSN),
the Jet Propulsion Laboratory (JPL) has developed a fibre link to
distribute reference signals from an H-Maser to synchronize each
antenna of the DSN \cite{LOGAN_LINK_APPLICATIONS},
\cite{CALHOUN}. \\
Modern cold atoms frequency standards in the microwave domain have
already demonstrated an accuracy in the 10$^{-15}$ range with the
potential to reach the 10$^{-16}$ level or better. Frequency
stabilities, defined by the Allan standard Deviation (ADEV), are
commonly of 10$^{-13}$ $\tau^{- \frac{1}{2}}$ for such standards
and a few 10$^{-14}$ $\tau^{- \frac{1}{2}}$ have  been
demonstrated using more advanced techniques \cite{FO2}. Cold atom
optical clocks have the potential to reach the 10$^{-17}$ accuracy
level \cite{Sr_SYRTE}, \cite{Ca_PTB}, \cite{Hg_NIST},
\cite{Sr_NPL}. The emergence of modern microwave-to-optical
synthesizers based on mode-locked femtosecond lasers allows high
resolution comparisons between microwave and optical clocks
\cite{FEMTO_NIST}, \cite{FEMTO_2}, \cite{LPL}. Clocks comparisons
are currently performed by satellite, as for example GPS or TWSTFT
(Two-Way Satellite Time and Frequency Transfer. Measurements are
limited by the transmission system to about 10$^{-15}$ at one day
averaging time \cite{CLOCK_COMPARISON}. Theses methods are thus
insufficient for measuring the ultimate performance of a microwave
or an optical standard (Fig. \ref{specification_100MHz.EPS}).

Upgrades of the orbital equipments are expectable to improve the
current performance, but are quite complex and expensive.
Moreover, the two previous systems deliver only a synchronization
signal not allowing direct short-term stability comparisons. Then
for much of applications a reference signal is needed. Hence, the
opportunity to compare microwave and optical clocks by the
development of a new type of a ground frequency dissemination by
optical fibre seems appropriate, even when the laboratories are
separated by 100 km \cite{BRAN}, \cite {NIST}, \cite{SATO}. One
can indeed take advantage of both the low attenuation and low
dispersion in the fibre, which allow reaching long distance
frequency transfer by maintaining a good signal-to-noise
ratio (SNR).\\
Moreover the access to an ultra-stable frequency reference for a
large number of laboratories open the way to perform new
experiments in fundamental physics. The development and operation
of a state-of-the-art frequency standard remain a strong
limitation and can be overcome by a fibre distribution system
connecting Time and Frequency Metrology laboratories to
users. \\
The simplest way to develop a fibre distribution is to use the
redundancy of the telecom network. In this paper, we present the
transfer of high frequency stability signal at 100 MHz, by using
the existing telecommunication fibre network, over a few tens
kilometers, with compensation of the phase noise introduced by the
link.
\section{Principle and objective}
The goal of the dissemination is the distribution of a reference
signal at a frequency of 100 MHz, synthesized from a frequency
standard, by amplitude modulation of an optical carrier, without
degradation of the phase noise of the distributed signal. The
reference signal modulates the bias current of a DFB laser diode,
at 1.55 $\mu$m, which is transmitted through a fibre optical link
to users. At the link extremity, a photodiode detects the
amplitude modulation and converts the optical signal to a
radio-frequency signal oscillating at the reference frequency and
phase coherent with the microwave reference source. \\
The high stability and low phase noise of the transferred signal
are degraded by the residual phase noise of the optical link and
by the attenuation in the fibre. We operate in urban environment
by using the existing telecom network. Thus, fibre layout and
installation aspects are not ideal and the stability of the
optical link can be affected by environmental effects. Optical
length of the fibre is modified by mechanical stresses and
temperature fluctuations. The first one affects phase noise and
short-term frequency stability performances of the transmitted
signal. The second effect, is a slowly changing
phenomenon and has an impact on the long-term stability. \\
These instabilities have been studied on two optical links using
the dense France Telecom network and connecting LNE-SYRTE to
Laboratoire de Physique des Lasers (LPL) (about 43 km), and
LNE-SYRTE with Laboratoire Kastler Brossel (LKB - University
Paris VI) (about 3 km).\\

\begin{figure}[!hbt]
  \centering
  \includegraphics[width=\columnwidth]{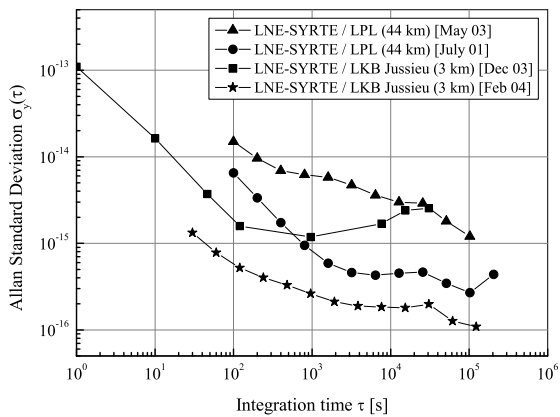}\\
  \caption{Frequency stability measurements of the LNE-SYRTE/LPL and LNE-SYRTE/LKB optical links}
  \label{links_stabilities.eps}
\end{figure}

Measurements, realized at different periods, are presented in
figure \ref{links_stabilities.eps} and show non-stationary effects
depending on the activities around the link. Periodic effects as
daily temperature variations appears as a bump at the half period,
on the ADEV. The frequency instabilities related to a sinusoidal
temperature perturbations can be calculated from the equation
(\ref{ADEV for sinusoidal variation}):
\begin{equation}\label{ADEV for sinusoidal variation}
    \sigma_y(\tau) = \Delta T \times \frac{TCD \times n \times
    L}{c} \times \frac{\sin^2(\pi \tau \nu_0)}{\tau}
\end{equation}
with $\Delta T$ the amplitude of the temperature fluctuation
[$^\circ$C], $TCD$ the thermal coefficient of delay
[ppm/$^\circ$C] of the optical fibre (typically 7 ppm/$^\circ$C
for standard telecom SMF28 fibre), $n$ the fibre core index, $L$
the optical link length [km], $c$ the light velocity in vacuum [3
$\times$ 10$^8$ m/s], $\nu_0$ the perturbation frequency [Hz], and
$\tau$ the averaging time [s]. For example, if we consider a
sinusoidal perturbation of 0.2$^{\circ}$C with a period of 1000s
due to air conditioning and acting on a section of 50 meters of
the optical link, the ADEV of the link could be limited to about
7x10$^{-16}$ at 500 s integration time. In the same way, a daily
0.5$^{\circ}$C temperature variation on 43 kilometers of optical
fibre is converted into an instability of the order of
1.3x10$^{-14}$ at 43200 s averaging time.\\
Consequently, the distribution system needs an active control loop
to compensate for these phase variations induced on the signal
transmitted through the link related to the environment
(mechanical vibrations, temperature fluctuations ...). \\
The objective of the dissemination being clock comparisons or
delivery of a reference signal coming from an H-Maser or a
Cryogenic Sapphire Oscillator (CSO), the compensation set-up must
introduce a phase noise lower than the reference signal. In this
perspective we have to develop a system which delivers a reference
signal at 100 MHz, showing a relative frequency stability
$\sigma_y(\tau) \leq 2.10^{-14}$ [$\tau$ = 1s] ($< 10^{-16}$ @
1d), that implies a residual flicker phase noise of -120
dBrad$^2$/Hz at 1 Hz and a white phase noise floor with a level of
-140 dBrad$^2$/Hz.

\section{Active phase fluctuations compensation system}

    \subsection{Presentation}
\begin{figure}[!hbt]
  \centering
  \includegraphics[width=\columnwidth]{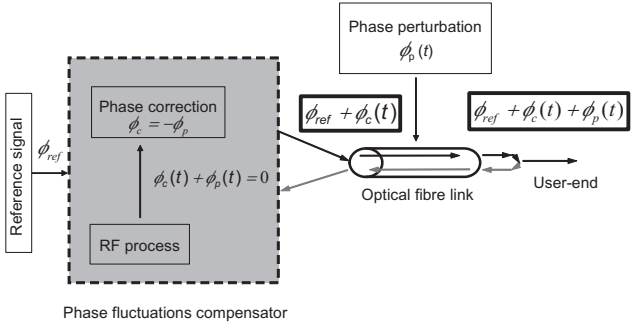}\\
  \caption{Schematic of the phase fluctuations compensation}\label{COMPENSATION_PRINCIPLE.eps}
\end{figure}

The principle of the phase fluctuations compensation, is displayed
in figure \ref{COMPENSATION_PRINCIPLE.eps}. At the link extremity,
the detected signal can not be directly compared to the reference
signal and thus the correction of the phase perturbations can be
only carried out at the link emission. A two-way distribution,
using the same optical fibre link, allows determination of the
phase perturbation accumulated along a full round trip with the
hypothesis that the forward and the backward signals are corrupted
by the same perturbation. The compensation rests then on the
measurement of the phase of the signal after one round trip to
apply a correction on the emitted signal. \\

The reference signal at the frequency f$_{\text{ref}}$ =
$\omega_{\text{ref}}/ 2 \pi$ is used for modulating a laser diode.
The amplitude modulated signal is then corrected by a phase term
$\phi_c$. This correction term is provided either by phase
shifting the RF modulating signal or by modifying the propagation
delay in the fibre. At the user-end, the signal corrupted by the
environmental perturbations is detected:
    \begin{equation}
    V_{\text{RF \ detected}} (t) \propto \sin (\omega_{\text{ref}} \ t + \phi_{\text{ref}} + \phi_c + \phi_p)
    \end{equation}
This signal is split in two signals: one part for the user
applications and the other to be re-injected via an optical
circulator in the same optical fibre. After one round-trip, the
signal, twice corrupted by the term $\phi_p$ is detected. A RF
process allows generation of an error signal, applied to the phase
corrector. Two different laser sources, operating at slightly
different wavelengths, are used for generating the forward and the
backward optical signals and optical add/drop functions are
realized with
optical circulators. \\
Different approaches of phase compensation have been studied and
are described here.

    \subsection{Electronic phase fluctuations compensator}

\begin{figure}[!hbt]
  \centering
  \includegraphics[width=\columnwidth]{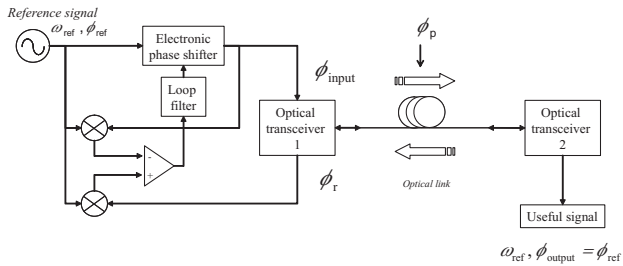}\\
  \caption{Simplified schematic of the phase conjugator}\label{simplified_phase_conjugator.eps}
\end{figure}

In the case of an electronic phase fluctuations compensator (cf
fig. \ref{simplified_phase_conjugator.eps}), the correction is
performed by acting on the phase of the injected signal in the
optical link, that we call $\phi_{\text{input}}$.

We define by $\phi_r$ the phase of the round-trip signal, and
$\phi_{\text{output}}$the phase of the detected signal at the
user-end, equal to:
\begin{equation}
    \phi_{\text{output}}(t) = \phi_{\text{input}} (t-\tau) +
    \int^t_{t-\tau} \phi_p(\xi) d\xi
\end{equation}
where $\tau$ is the propagation delay in the optical fibre link
and $\phi_p(\xi)$ is the distributed phase perturbation along the
fibre. The main effect of the delay $\tau$ is to limit the loop
bandwidth. In the following discussion, we neglect the influence of the delay.\\

\begin{figure}[!hbt]
  \centering
  \includegraphics[width=\columnwidth]{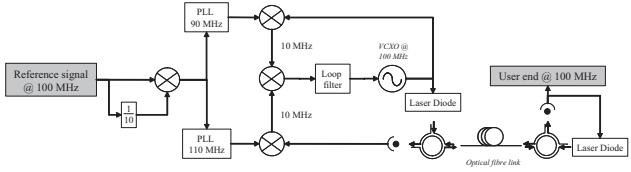}\\
  \caption{Block diagram of the full electronic compensation system}\label{ELECTRONIC_COMPENSATOR.eps}
\end{figure}

The output signal must be phase coherent with the reference source
of frequency $\omega_{\text{ref}}$ and on average of phase
$\phi_{ref}$, and thus the correction applied to the emitted
signal must be equal to the opposite of the phase perturbation
$\phi_p$. Consequently, on average (or for time much longer than
$\tau$) the phase of the input signal, $\phi_\text{input}$ is:
\begin{equation}
    \phi_{\text{input}} = \phi_{\text{ref}} - \phi_p
\end{equation}
Then, the phase of the round-trip signal becomes:
\begin{eqnarray*}
    \phi_{r} &=& \phi_{\text{input}} + 2 \times \phi_p \\
             &=& \phi_{\text{ref}} + \phi_p
\end{eqnarray*}
The phase coherence of the output signal is hence imposed by
maintaining a conjugation relationship between the input and the
round trip signal of the optical link:
\begin{equation}
     (\phi_{\text{input}} - \phi_{\text{ref}}) = - ( \phi_r - \phi_{\text{ref}})
\end{equation}
A simplified scheme of the phase conjugator is shown in figure
\ref{simplified_phase_conjugator.eps}. The correction is performed
with a phase shifter in series with the reference signal, which is
used as the input signal. The reference signal is power divided to
drive two phase detectors. Phase detection between the reference
signal, the input corrected signal and the round-trip signal,
allow generation of two baseband signals, connected to the inputs
of a low noise differential amplifier. The output of this
amplifier is used for driving a loop filter, controlling the
electronic phase shifter until the phase conjugation, and thus a
zero level at the amplifier output is reached. Although the
simplicity of operation, this system suffers from various
drawbacks. First, the phase correction is limited by the dynamic
of the phase shifter. Electronic phase shifters have a typical
dynamic of 180 degrees with a non linear response, inducing
variable insertion losses. Moreover the phase shifter can present
a phase noise excess, compared to the other components of the
phase conjugator. Secondly, phase detectors are quite sensitive to
the driving levels and it is difficult to ensure the same
sensitivity for the two detectors of figure
\ref{simplified_phase_conjugator.eps}. \\
The practical realization leads to a very poor effective system of
the phase perturbations cancellation. A new scheme, regarding the
previous considerations and introduced by the JPL \cite{CALHOUN2}
is shown in figure \ref{ELECTRONIC_COMPENSATOR.eps}. \\

Two symmetrical signals are produced by frequency shift
(f$_{\text{shift}}$) of the reference signal
(f$^\pm$=f$_{\text{ref}}$ $\pm$ f$_{\text{shift}}$). This scheme
allows replacement of the double phase measurements (Fig.
\ref{simplified_phase_conjugator.eps}) by a much more accurate
double
frequency mixing and a single phase measurement.\\
The dynamic and the linearity of the phase correction is improved
by using a voltage controlled quartz oscillator (VCXO), as a phase
shifter, delivering a signal at the reference frequency with a
stable amplitude. The VCXO presents thus the advantage to correct
all phase perturbation in the correction bandwidth of the phase
compensator, which is limited by the round-trip propagation delay
in the optical link (about 0.3 ms in the case of the 43-km
LNE-SYRTE to LPL optical link). \\
The 100 MHz output signal of the VCXO modulates the bias current
of the DFB laser diode. The optical signal is launched in the
optical fibre link to the user. At the user end, a simple system
allows detection and regeneration of the backward signal. The
detected signal after a one-way distribution is proportional to:
\begin{equation}
V_{\text{User end}}(t) \propto \sin(\omega_{\text{osc}} \times t +
\phi_{\text{osc}} + \phi_p)
\end{equation}
The backward optical signal is submitted to the same phase
perturbation and after one complete round-trip, the detected
signal has the following form:
\begin{equation}
V_{\text{round trip}} (t) \propto \sin(\omega_{\text{osc}} \times
t + \phi_{\text{osc}} + 2 \times \phi_p)
\end{equation}
The servo loop forces the VCXO at 100 MHz both to be phase
coherent with the reference source and to compensate for the phase
perturbation. For obtaining the phase conjugation, two signals
separated by 10 MHz around the reference frequency (one at 90 MHz
and the other at 110 MHz) are produced by frequency mixing between
the reference signal and itself frequency divided by ten. Two
different systems, based on PLL (Phase Lock Loop) are used for
filtering each signal issue from the previous frequency mixing.
The signal, from the "down conversion", at 90 MHz, is mixed with
the modulating signal, delivered by the VCXO, to obtain a signal
at 10 MHz:
\begin{equation}
V_1 (t) \propto \sin((\omega_{\text{osc}}- 2 \ \pi \times \text{90
MHz}) \times t + \phi_{\text{osc}} - \frac{9}{10}
\phi_{\text{ref}})
\end{equation}
In parallel, the signal at 110 MHz is mixed with the round-trip
signal, producing another signal at 10 MHz:
\begin{equation}
V_2 (t) \propto \sin((2 \ \pi \times \text{110 MHz} -
\omega_{\text{osc}}) \times t + \frac{11}{10} \phi_{\text{ref}} -
\phi_{\text{osc}}- 2 \phi_p)
\end{equation}
The phase comparison at 10 MHz allows generation of a base-band
signal, containing the three phase terms:
\begin{equation}
V_{\text{error}} (t) \propto \phi_{\text{osc}} + \phi_p -
\phi_{\text{ref}}
\end{equation}
which is cancelled in normal operation. The phase of the VCXO is
then:
\begin{equation}
\phi_{\text{osc}} = \phi_{\text{ref}} - \phi_p
\end{equation}
By this process, the stability and the accuracy of the reference
source is transmitted to the user end in the system bandwidth. \\
The capacity of the phase compensator to reject phase
perturbations in the control bandwidth is defined by the rejection
factor, equal to the ratio between the phase variations in open
and in closed loop. The performance of the distribution system
depend both of the intrinsic system phase noise and of the
rejection factor.\\

\begin{figure}[!hbt]
  \centering
  \includegraphics[width=\columnwidth]{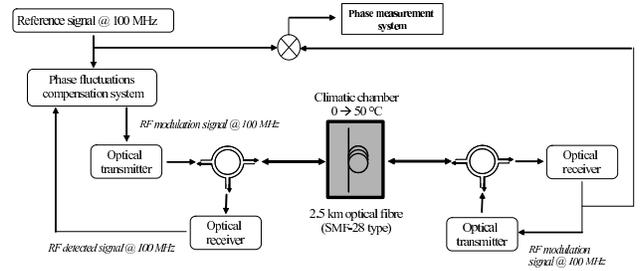}\\
  \caption{Block diagram of the compensation system test bench}\label{ELECTRONIC_COMPENSATION_TEST_BENCH.EPS}
\end{figure}

\begin{figure}[!hbt]
  \centering
  \includegraphics[width=\columnwidth]{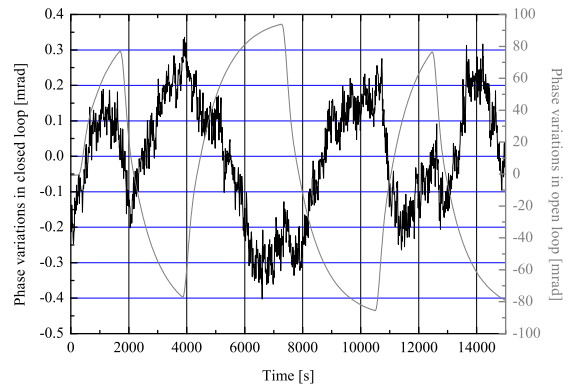}\\
  \caption{Phase shift induced by temperature modulation of the transmitted signal,
in open and closed loop at 100
MHz}\label{PHASE_vs_TEMPERATURE_ELECTRONIC_COMPENSATOR.EPS}
\end{figure}

Figure \ref{ELECTRONIC_COMPENSATION_TEST_BENCH.EPS} displays the
set-up used for the characterization of the phase conjugator.
Simulation of phase perturbations are realized by periodically
heating a 2.5-km fibre spool with an amplitude of 4$^\circ$C and a
period of about 4000 s. This perturbation induces a phase
modulation of the order of 200 mrad on the 100-MHz transmitted
signal. In operation, when the phase conjugator is activated, the
residual phase modulation measured at the link output is reduced
to 0.4 mrad (cf. figure
\ref{PHASE_vs_TEMPERATURE_ELECTRONIC_COMPENSATOR.EPS}), that
implies a rejection factor of the phase perturbations along
the link of about 500 (53 dB). \\

\begin{figure}[!hbt]
  \centering
  \includegraphics[width=\columnwidth]{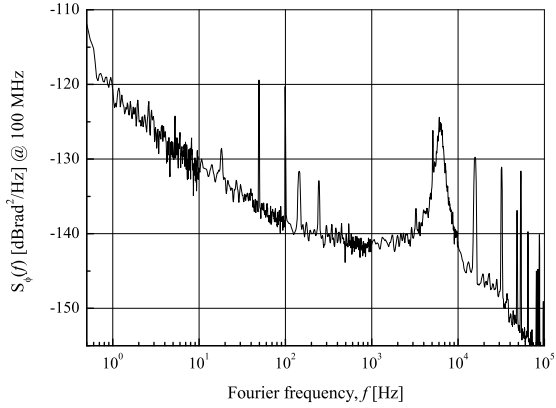}\\
  \caption{Phase noise spectral density of the full electronic compensation system (The phase noise bump comes from a gain excess
of the loop)}\label{PHASE_NOISE_ELECTRONIC_COMPENSATION.EPS}
\end{figure}

Moreover, the complete distribution system including the phase
conjugator shows a flicker phase noise with a level of -123
dBrad$^2$ at 1 Hz and a white noise floor below -140 dBrad$^2$/Hz
(Fig. \ref{PHASE_NOISE_ELECTRONIC_COMPENSATION.EPS}). This ensures
the possibility to transfer metrological signal with a frequency
stability $\sigma_y(\tau)$ below than 1$\times$10$^{-14}$ at 1
second
averaging time. \\

This system was implemented at SYRTE.

    \subsection{Optical compensation system}
\begin{figure}[!hbt]
  \centering
  \includegraphics[width=\columnwidth]{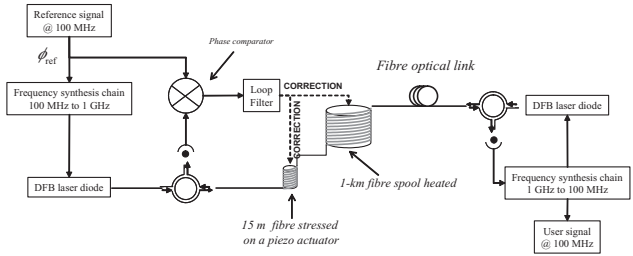}\\
  \caption{Schematic of the optoelectronic system for phase fluctuations compensation}\label{OPTOELECTRONIC_COMPENSATOR.eps}
\end{figure}

An optoelectronic compensation system has also been developed and
implemented at LPL, operating in a slightly different way. The
phase correction is applied both on the emitted and on the
backward signal by directly acting on a section of optical fibre,
placed in series with the optical link. The phase correction is
then performed by modifying the optical propagation delay (and
thus the optical path) of the optical signal in the fibre link.
The principle of the optoelectronic phase compensation is
presented in figure \ref{OPTOELECTRONIC_COMPENSATOR.eps}.

\begin{figure}[!hbt]
  \centering
  \includegraphics[width=\columnwidth]{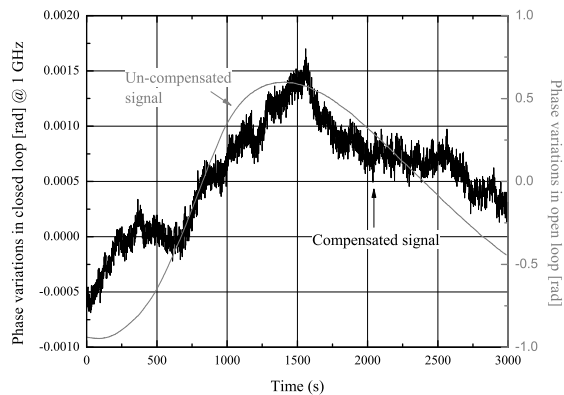}\\
  \caption{Phase variations measurement, induced by a 10 $^\circ$C temperature perturbation on a 1-km standard fibre link, in open and closed loop}\label{PHASE_vs_TEMPERATURE_OPTICAL_COMPENSATOR.EPS}
\end{figure}

For increasing isolation between the two ways of the distribution
system, two different modulation frequencies are used: 1 GHz for
the emission and 100 MHz for the return way. Consequently, a 1-GHz
signal must be generated from the reference 100-MHz signal by a
frequency synthesizer. The optical signal is distributed in the
fibre link through the optical compensator and thus corrected. At
the user-end, the detected signal has the following form:
\begin{equation}
V_{\text{user}} \propto \sin(2 \ \pi \times \text{1 GHz} \times t
+ 10 \times \phi_{\text{ref}} + \phi_{\text{correction}}' +
\phi_p')
\end{equation}
where $\phi_{\text{correction}}'$ and $\phi_p'$ are respectively
the correction term and the perturbation applied to the
transmitted signal at 1 GHz.\\
A second frequency synthesis chain allows delivering a signal at
100 MHz from the 1-GHz detected signal. This signal is used for
modulating a second laser diode to produce the backward optical
signal. This signal is corrupted by the same perturbation and also
corrected. After one round trip, we detect a 100 MHz signal
proportional to:
\begin{equation}
V_{\text{round trip}} \propto \sin(2 \ \pi \times \text{100 MHz}
\times t + \phi_{\text{ref}} + 2 \times (\phi_{\text{correction}}
+ \phi_p))
\end{equation}
with:
\begin{eqnarray*}
  \phi_\text{correction} &=& \frac{\phi_\text{correction}'}{10} \ \text{and} \\
  \phi_p &=& \frac{\phi_p'}{10}
\end{eqnarray*}
The phase comparison between the reference signal at 100 MHz and
the round-trip signal leads to generation of a baseband error
signal:
\begin{equation}
V_{\text{error}} \propto \phi_{\text{correction}} + \phi_p
\end{equation}
which is applied to a loop filter to drive the optical phase
corrector. This corrector is composed of two sub-systems. Fast and
small phase fluctuations (mechanical vibrations, fast temperature
variations induced by air conditioning eg.) are corrected by
changing the length of a portion of optical fibre with a
piezo-electric actuator. A 15-meter optical fibre is wrapped
around a 5-cm diameter PZT of about 10 $\mu$m variation under 1 kV
voltage, and is mechanically stretched, allowing correction up to
about 15 ps or 10 mrad at 100 MHz, in a bandwidth of a few
hundreds Hz. \\
Slow and large perturbation are compensated by heating a 1-km
fibre spool introduced along the link ($\simeq$
40 ps/$^\circ$C or 25 mrad/$^\circ$C at 100 MHz).\\
For the laboratory tests, a fibre stretcher and an heating system
are placed along a laboratory link of a few kilometers to generate
fast and slow phase perturbations. By this way, a temperature step
of 10 $^{\circ}$C is realized on a 1-km fibre spool, and the phase
shift induced on the detected signal and measured in open and
closed loop is reported in Fig.
\ref{PHASE_vs_TEMPERATURE_OPTICAL_COMPENSATOR.EPS}. Rejection
factor of about 750 for slow phase perturbations is shown.

\begin{figure}[!hbt]
  \centering
  \includegraphics[width=\columnwidth]{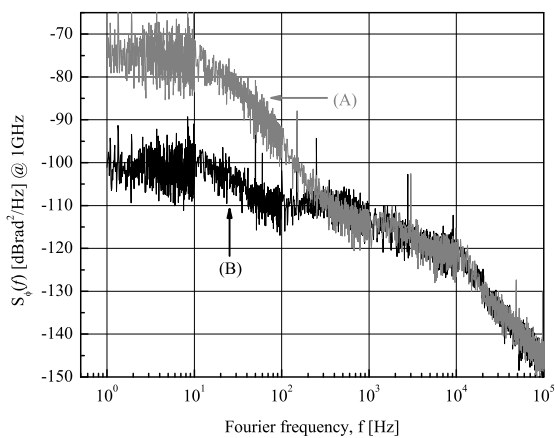}\\
  \caption{Measurements of the PZT phase noise reduction: Open
loop phase noise measurement intentionally degraded (A) to show
the rejection due to the PZT corrector
(B)}\label{fast_phase_fluctuations_compensation.EPS}
\end{figure}

In contrast, only 20-25 dB attenuation (Fig.
\ref{fast_phase_fluctuations_compensation.EPS}) are observed on
intentionally produced small and fast perturbations. The gain of
the correction is limited by parasitic phase shifts generated by
Polarization Dependent Losses (PDL) under mechanical stress of the
fibre. The mechanical stress affects the geometry of the fibre
which becomes birefringent. Thus the polarization of the
transmitted beam changes and leads to an amplitude modulation (AM)
of the detected signal on the tilted photodiode. AM is directly
converted into PM (Phase modulation) and detected as a phase
perturbation term which is compensated by the phase compensator.
Corrections of phase perturbations are corrupted by this parasitic
phenomena and the performance of the system is then degraded. PDL
is only related to the optical signal and independent of the
modulation frequency. By upgrading the system to higher operation
modulation frequencies, this problem
could be reduced proportionally to the frequency.\\

    \section{Characterization of the two different systems in a full directional link}
        \subsection{Set up}
A full bidirectional compensated optical link (2$\times$43 km) has
been achieved by using the two fibres of the LNE-SYRTE to LPL link
and by implementing the two previous compensation systems. This
link is composed of various sections of buried optical cables of
the France Telecom metropolitan network. The continuity of each
optical fibre of 43 km is ensured by optical splicing and a global
attenuation of 12 dB on each fibre is measured.\\  The low phase
noise 100-MHz local oscillator of LNE-SYRTE is transferred to LPL
by using one of the two fibres of the link, and is phase
compensated by the phase conjugator. At LPL, a signal phase
coherent with the LNE-SYRTE local oscillator is detected and used
as the input reference signal for the second optical link,
connecting back LPL to LNE-SYRTE via the second 43-km fibre. The
optoelectronic system is installed on this
link to compensate for the phase perturbations. \\
At LNE-SYRTE, we compare the phase detected signal coming from LPL
with the local oscillator for characterizing the two dissemination
systems. The use of two different systems allows systematic
studies related to one of the two compensators, and thus allows to
have a full characterization
of the distribution system. \\

\begin{figure}[!hbt]
  \centering
  \includegraphics[width=\columnwidth]{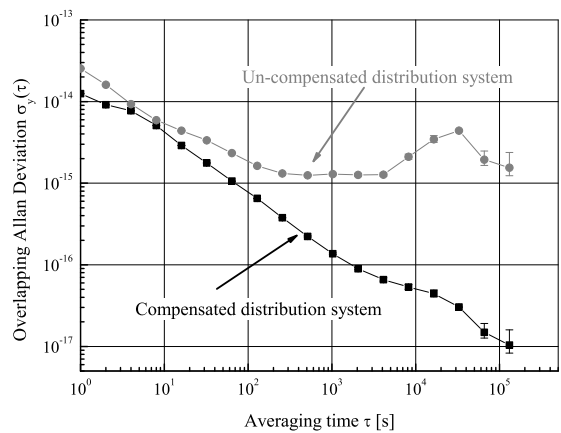}\\
  \caption{Frequency stability of the complete 86-km distribution in open and closed loop}\label{LINK_88KM.EPS}
\end{figure}

Figure \ref{LINK_88KM.EPS} shows frequency stability measurements
of the full bidirectional link (2$\times$43 km) between LNE-SYRTE
and LPL. A fractional frequency stability of 10$^{-17}$ at one day
and 1.2$\times$10$^{-14}$ at one second averaging time is
obtained. These values of ADEV are calculated from the phase data
measured on the link and filtered with a low-pass filter of 3 Hz.

    \subsection{Direct application of the LNE-SYRTE to LPL optical link}
This link has been used to compare an optical frequency standard
against an ultra-stable microwave oscillator, separated
by a 43 km fibre link. \\
The optical standard is a CO$_\text{2}$ laser stabilized on an
OsO$_\text{4}$ molecular absorption operating in the infrared
domain, at 30 THz \cite{BERNARD}. At LNE-SYRTE, a 100-MHz signal
is synthezised from the signal of a Cryogenic Sapphire Oscillator
at 12 GHz, weakly phase locked on the signal of an H-MASER
\cite{CHAMBON}.\\
A femtosecond laser frequency comb allows the optical to microwave
comparison and we demonstrate a resolution of 3$\times$10$^{-14}$
at one second integration time \cite{DAUSSY}.

    \subsection{Systems limitations}

As shown in figure \ref{LINK_88KM.EPS}, the frequency stability in
closed loop is better than the open loop for all integration
times, between 1 s and 1 day. \\
The short-term stability is mainly limited by the SNR at the
detection, degraded by three contributions: the intrinsic noise of
the distribution system, the optical attenuation in the fibre and
the stimulated Brillouin scattering (SBS). Optical losses can be
overcome by injecting powerful signals in the fibre, up to the
Brillouin generation in order to avoid any laser source depletion.
The direct amplitude modulation of the DFB bias current induces a
broadening of the optical spectrum with a distribution of the
energy within this spectrum and thus allows increasing the
injected power level. The SBS leads also to a scattered wave
detected as a white phase noise excess around the RF carrier with
a few tens MHz width. This can be easily reduced by additional
fibre Bragg grating filters (FBG), used in detection. The
optoelectronic phase compensator, operating with two different
modulation frequencies, separated by more than the Brillouin
bandwidth, is less sensitive to this phenomenon. \\
In spite of its complexity, the phase conjugator presents a better
phase noise ensuring thus a better short-term performance. The
noise of the optoelectronic system is mainly degraded by additive
amplitude noise, generated by the PZT corrector. One solution for
improving the phase noise of each dissemination system is then to
move to a higher operation RF frequency as 1 GHz. \\
The long-term frequency stability is mainly limited by the phase
conjugator which presents a rejection factor of only a few
hundreds. Two main phenomena degrade the performance of the phase
conjugator. Coherent leakage signals of 90 MHz and 110 MHz may
induce parasitic phase shifts. The second limitation comes from
optical feedback due to reflections of optical connectors or
optical splicing along the link. A parasitic signal with an
undetermined phase, varying in time, is thus detected and may be
non-negligible compared to the main detected signal. At the
detection, the ratio between the main signal and the parasitic
signal is proportional to:
\begin{eqnarray*}
\frac{10^{- \alpha (L - 2 L_R)}}{R}
\end{eqnarray*}
where $\alpha$ is the optical attenuation in the fibre [dB/km],
$R$ is the power reflection coefficient, and $L_R$ [km] and $L$
[km] are
respectively the reflection distance and the link lengh.\\
To ensure a sufficient compensation of the phase fluctuations
introduced by the link, all parasitic noises should be 60 dB under
the the detected signal. Such level could be reached by shifting
the modulation frequency of the backward signal or by adding
optical filters. \\
Finally, we are also confronted by a polarization effect, PMD
(Polarization Mode Dispersion), which are detected as a chromatic
dispersion with a random coefficient, leading to a random
propagation delay on each way of the dissemination. Due to PMD,
the principle of the compensation, based on the measurement of
twice the perturbation after one round trip is not valid any more.
One solution is the polarization scrambling of the injected
optical signal, faster than the loop bandwidth.

\section{Conclusion $\&$ perspectives}
We have demonstrated for the first time the long-distance transfer
of both short-term and long-term frequency stability of frequency
standards, with low phase noise via telecom optical fibres. A
stability of a few 10$^{-14}$ at one second and 10$^{-17}$ for one
day integration has been obtained on an optical link of 86 km. \\
In this way, we are able to compare two distant frequency
standards operating in the microwave domain (CSO) and in the
infrared domain (stabilized CO$_2$ laser) with a resolution of
3$\times$10$^{-14}$ at 1 s. \\
The electronic phase conjugator presents the advantage to have an
infinite dynamic allowing compensation of all phase perturbations,
with a bandwidth limited by the round trip delay in the optical
link. With a better phase noise, this setup shows a good rejection
factor for the short-term but is limited to a few hundreds for the
long-term. Even if, the user end of the dissemination system is
really simple, the system remains complex. In contrast, despite
its relative simplicity, the optoelectronic compensator shows a
limited short-term rejection factor due to PMD and PDL, but a
better long-term rejection, than the one achieved with the
electronic compensation setup.\\
We intend to improve these results by one order of magnitude by
upgrading both dissemination setups. Moving to a modulation
frequency of 1 GHz should improve the intrinsic noise of each
system. Additional polarization scramblers should help to reduce
polarization effects and amplitude noise caused by the PZT
corrector. We also plan to modify the phase conjugator by
frequency shifting the backward signal to suppress the effect of
the optical feed-back and high-order products in frequency mixers,
and to use of optical filters at the same time.
\section{Acknowledgments}
This work was supported by the European Space Agency / ESOC.\\
Thanks to Observatoire de Paris, the Laboratoire National de
métrologie et d'Essais (LNE), Paris 13 University, the French
Research Ministry, CNRS, and the
Laboratoire d$'$Optronique, GIS FOTON, ENSSAT Lannion.\\
The authors thank D. Chambon and L. Volodimer.
\newpage

\section*{List of References}

\end{document}